\documentclass[]{PoS}
\def\lsim{\mathrel{\lower3pt\hbox{$\sim$}}\hskip-11.5pt\raise3pt\hbox{$<$}\;}

\title{A light scalar WIMP through the Higgs portal?}

\ShortTitle{A light scalar WIMP through the Higgs portal?}

\author{\speaker{Michel H.G. Tytgat}
\\
        Service de Physique Th\'eorique, CP225\\
        Universit\'e Libre de Bruxelles\\
        Bld du Triomphe, 1050 Brussels, Belgium\\
        E-mail: \email{mtytgat@ulb.ac.be}}

\abstract{In these proceedings, I report on the status of a simple singlet scalar dark matter model in the light of recent results from both direct detection experiments, in particular DAMA, CoGeNT, CDMS-II and Xenon10/100, and indirect searches, in particular {\it Fermi}-LAT. Specifically, I confront the light scalar WIMP candidates, $M_{DM} \sim $ few GeV, that are consistent with CoGeNT and/or DAMA, to constraints that may be set using the recent {\it Fermi}-LAT data on Milky Way dwarf spheroidal galaxies (dSphs) and the isotropic diffuse gamma-ray emission. I show that the latter observations set relevant exclusion limits on the lightest WIMP candidates.}

\FullConference{Identification of Dark Matter 2010-IDM2010\\
		July 26-30, 2010\\
		Montpellier France}

\begin{document}


There is much interest recently in light dark matter (DM) candidates, by which I mean here WIMPs\footnote{By WIMP I mean here a neutral, stable particle with a relic abundance fixed by thermal freeze-out, {\em i.e.} $\langle \sigma v\rangle \sim 10^{-26}$ cm$^3\cdot$s$^{-1}$. This does not imply that the candidate has electroweak interactions, or has a mass related to the electroweak scale, and thus potentially includes many scenarios, including WIMPless particles \cite{Kumar:2010qc}.} with a mass in the GeV range. As everybody at this conference knows, this is mostly due to the recent (and less recent) hints provided by some direct detection experiments. In particular, last March the CoGeNT collaboration released an analysis of their first round of data taking, which revealed an excess of events at low recoil energies \cite{Aalseth:2010vx}. If interpreted as being due to Spin-Independent (SI) elastic scattering of a DM particle, the excess points to a candidate with a mass in the GeV range, and a SI cross-section $\sigma_n^0 \sim 10^{-40}$ cm$^2$. Intriguingly, this range of parameters is not very different (but not altogether consistent) with that required to explain the annual modulation observed by the  DAMA/LIBRA and DAMA/NaI (DAMA for brief) experiments \cite{Bernabei:2008yi}.\footnote{A major problem with such an interpretation is that the unmodulated rate must be also  large, which implies that the background must be anomalously low in the region of recoil energies where the signal is observed by DAMA, see for instance \cite{Fairbairn:2008gz}.} There also been very tentative hints from the CRESST experiment (unpublished, but discussed  by W. Seidel at this conference \cite{Seidel}), and there also the few events observed by CDMS-II \cite{Ahmed:2009zw}, which are both consistent with a light DM candidate. In the sequel, I will refer collectively  to these results as CoGeNT-DAMA. Of course, there are also strong direct constraints on such a light candidate, most notably from the CDMS-Si \cite{Akerib:2005kh}, and the Xenon10 \cite{Angle:2008we,Angle:2009xb} and Xenon 100 \cite{Aprile:2010um} collaborations. Whether the signals and exclusion limits may be consistent with each others is still being debated \cite{Hooper:2010uy}, but I believe it is fair to say that {\it \c ca sent le roussi} for CoGeNT-DAMA, specially in the light of two new results. The first is the analysis of liquid Xe experimental data proposed by P. Sorensen at this conference, which is less sensitive to experimental uncertainties, like scintillation efficiency, and quite efficiently at constraining light DM candidates \cite{Sorensen}. The other is the recent re-analysis of CDMS data, with a focus on low recoil energies \cite{Ahmed:2010wy}. Both studies exclude the CoGeNT-DAMA candidates. However mourning the CoGeNT-DAMA results is perhaps premature and, besides, I believe that it will remain of interest in the future to study the possibility of a light DM, both from the theoretical and experimental point of views.

There are in the literature two complementary approaches to the study of potential light WIMP candidates, with both pros and cons. The model-independent approach, which is based on effective operators, each of which may encompass many models, depends by construction on very few parameters but has generally limited predictive power (see {\em e.g.} \cite{Kopp:2009qt,Fitzpatrick:2010em,Chang:2010yk,Essig:2010ye}). The model dependent approach is generally limited in scope, but has potentially a richer phenomenology (see {\em e.g.} \cite{Barger:2010yn,Belikov:2010yi,Mambrini:2010dq}). Here I briefly review the status of the simplest real singlet scalar model \cite{Silveira:1985rk,McDonald:1993ex,Burgess:2000yq,Andreas:2010dz}. The salient features of this toy model are the following. The stability of the scalar, $S$,  is ensured by a $Z_2$ symmetry, introduced by hand, and not broken in the vacuum. The $S$ interacts with the Standard Model particles only through the Standard Model Higgs (hence the Higgs portal). This implies that elastic scattering with nuclei (like in direct detection) and annihilation (like in the early universe) are related \cite{Andreas:2008xy}. To put differently, there is just one family of relevant effective operators at low energies in this model,
$$
{\cal O}_S \sim {\lambda_S m_f\over M_h^2}  S^2 \bar f f,
$$
where $\lambda_S$ is the $S$-Higgs quartic coupling, and $M_h$ is the Standard Model Higgs mass (and thus only the combination $\lambda_S/M_h^2$ is relevant) and $m_f$ is a fermion mass. The mass of the scalar DM is given by
$$
M_S^2 = \mu_S^2 + \lambda_S v^2
$$
with $v=246$ GeV, and  $\mu_S$ is the $S$ bare mass. I remark that this expression implies that a light $M_S$ is not natural in this scheme, as it requires a cancellation between the two terms on the RHS ({\em i.e.} fine-tuning). The important outcome of these basic properties is that the cosmic relic abundance is determined if agreement with CoGeNT-DAMA is imposed. It is unique properties of such a scalar candidate, interacting through a scalar particle ({\em i.e.} the Higgs), that its annihilation cross-section is then predicted to be $\langle \sigma v\rangle ={\cal O}(3 \cdot 10^{-26}$ cm$^3\cdot$ s$^{-1})$ (see Figure I, and also, for instance \cite{Fitzpatrick:2010em}, for other SM candidates and interaction channels).

The model has little freedom left and may be rather easely falsified. In the sequel we will consider constraints from indirect detection but before this we briefly mention one striking consequence of the model for the search of the Higgs at the LHC. Agreement with CoGeNT-DAMA requires the $S$ to be rather light, and to have a substantial coupling to the Standard Model Higgs, $\lambda_S = {\cal O}(0.1)$. These imply that the branching ratio for the Higgs to decay into an $S-S$ pair is close to one, at least for a Higgs below the threshold for massive gauge boson production \cite{Andreas:2008xy}. 

Within the present framework, the annihilation of the $S$ into various messengers (gamma-rays, positrons, anti-protons, neutrinos) is uniquely predicted and, modulo the usual astrophysical uncertainties, may be used to constrain further the model and its siblings. Here we consider two specific signatures: the annihilation into gamma-rays in Milky Way dSphs, and  the isotropic diffuse gamma-ray emission, both in the light of recent {\it Fermi}-LAT data. 

Due to their large mass-to-light ratio, and their limited astrophysical gamma-ray activity, dSphs are potentially good places to search for gamma-ray signatures of DM, despite the expected small statistic. The {\it Fermi}-LAT collaboration has studied 14 nearby dSphs within the Milky Way. The first 11 months data revealed no gamma-ray excess and thus exclusion limits have been set on the flux of gamma-rays above 100 MeV from these dSphs, for candidate heavier than 10 GeV \cite{Abdo:2010ex}. Assuming a NFW profile (properly adapted to suit dSphs), the collaboration has also set limits on the annihilation cross-section of specific dark matter models, most notably the neutralino and Kaluza-Klein DM, with mass $M_{DM}$ above 30 GeV, and candidates annihilating into $\mu^+-\mu^-$ (relevant for the PAMELA excess) above 10 GeV. In \cite{Andreas:2010dz} we have used these data to set limits on the flux from $S$ candidates. Here we give in Table 1 the limits on the total annihilation cross section. Strictly speaking, the limit may only be set for a candidate with $M_S \geq 10$ GeV, so the limit on the 6 GeV and 8 GeV candidates are based on a naive extrapolation. At this conference, Llena Garde, from the {\it Fermi}-LAT collaboration, has reported  preliminary results from a stacking analysis of 8 dSphs (for 22 months of data) which gives strong constraints and this down to $M_{DM} = 5$ GeV, potentially excluding candidates below 30-40 GeV, for annihilation into $\bar b-b$ \cite{Llenagarde}. The constraints for annihilation into $\tau^+\tau^-$ may be expected to be similar. As an instance we give limits for an 8 GeV candidate annihilating into $\tau^+\tau^-$, as advocated in the recent discussions around  gamma-ray emission excess at the galactic centre \cite{Hooper:2010mq}. More statistics would give more stringent constraints but it is as likely that a cored  DM profile in dSphs (instead of a cuspy NFW) would relax the tension, if any. 
\begin{table}[t]
  \begin{center}
    \begin{tabular}{c|c|c|c}
$M_{DM}$ & BR & {Ursa Minor} & {Draco}  \\
\hline\hline
      10 GeV &
       BR($SS \rightarrow \tau^-\tau^+$) $\simeq 10\%$ & $\leq 2.6$ & $\leq 2.9$ \\
      & BR($SS \rightarrow b\bar{b}+ c\bar{c}$) $\simeq 90\%$ &  &\\ 
      \hline
       6 GeV &  
      BR($SS \rightarrow \tau^-\tau^+$) $\simeq 20\%$ & $\lsim 2$&$\lsim 2$ \\	
      &BR($SS \rightarrow b\bar{b}+ c\bar{c}$) $\simeq 80\%$ & & \\ 
\hline
8 GeV & 
BR($X X \rightarrow \tau^+\tau^-) = 100 \%$ & $\lsim 2.4 $ & $ \lsim 2.5$ \\
      \hline\hline
    \end{tabular}    
       \label{tab:dwarfs}
  \end{center}
\caption{95 C.L. exclusion limits on the annihilation cross-section ($\sigma v$ in units of $10^{26}$ cm$^3\cdot$s$^{-1}$) based on the limits on the flux of gamma-rays set by {\it Fermi}-LAT for two representative dSphs (Ursa Minor and Draco), using the median value of the J-factors \cite{Abdo:2010ex}. The last line is relevant for the 8 GeV candidate of Ref.\cite{Hooper:2010mq}}
\end{table}

The last topic that I would like to address is complementary constraints on light WIMPs (and the $S$ in particular) from the isotropic diffuse gamma-ray emission. This is again based on {\it Fermi}-LAT data, released in \cite{Abdo:2010dk}, which constrains candidates above $10$ GeV (see also \cite{Abazajian:2010sq}). We have done our own analysis, using distinct astrophysical assumptions regarding the distribution of dark matter halos as a function of redshift \cite{Arina:2010rb}. Our results are consistent with those of \cite{Abdo:2010dk,Abazajian:2010sq}, but our constraints extend to slightly lighter candidates, potentially excluding candidates which have not yet been probed by direct detection experiments (see Figure 2). The exclusion limits are of course quite sensitive to the astrophysical uncertainties (we have adopted a NFW profile for early DM halos). Most critical is the mass of the lightest halo that may be formed, a parameter that depends on the temperature of kinetic decoupling of the DM candidate (see {\em e.g.} \cite{Bringmann:2009vf}). For the case of the $S$, which interacts with Yukawa couplings to the Standard Model particles, the couplings to $e^+e^-$ and neutrinos is completely negligible, and kinetic decoupling occurs close to the chemical freeze-out, $T \sim 150$ MeV (for the sake of comparison, typical neutralino candidates decouple close to $T \sim 1$ MeV). This in turn implies that quite light (and thus dense) dark matter halos may form in the early universe, so that the constraints are comparatively stronger (see \cite{Arina:2010rb} for a detailed discussion). 
\begin{figure}[t!]
\begin{minipage}[t]{0.5\textwidth}
\centering
\includegraphics[width=0.9\columnwidth]{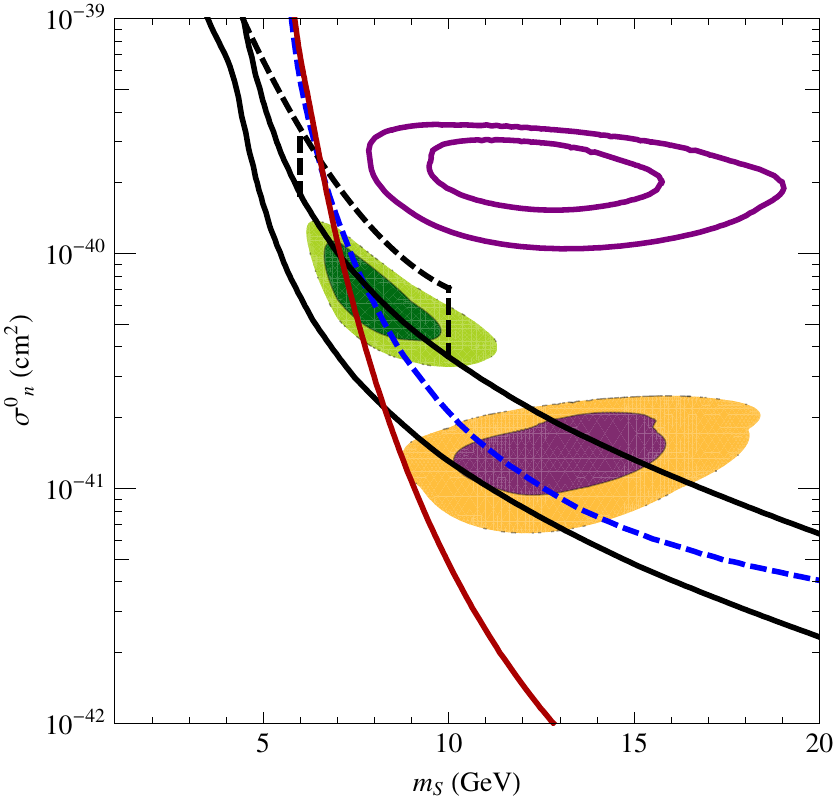}
\end{minipage}
\hspace*{-0.2cm}
\begin{minipage}[t]{0.5\textwidth}
\centering
\includegraphics[width=0.9\columnwidth]{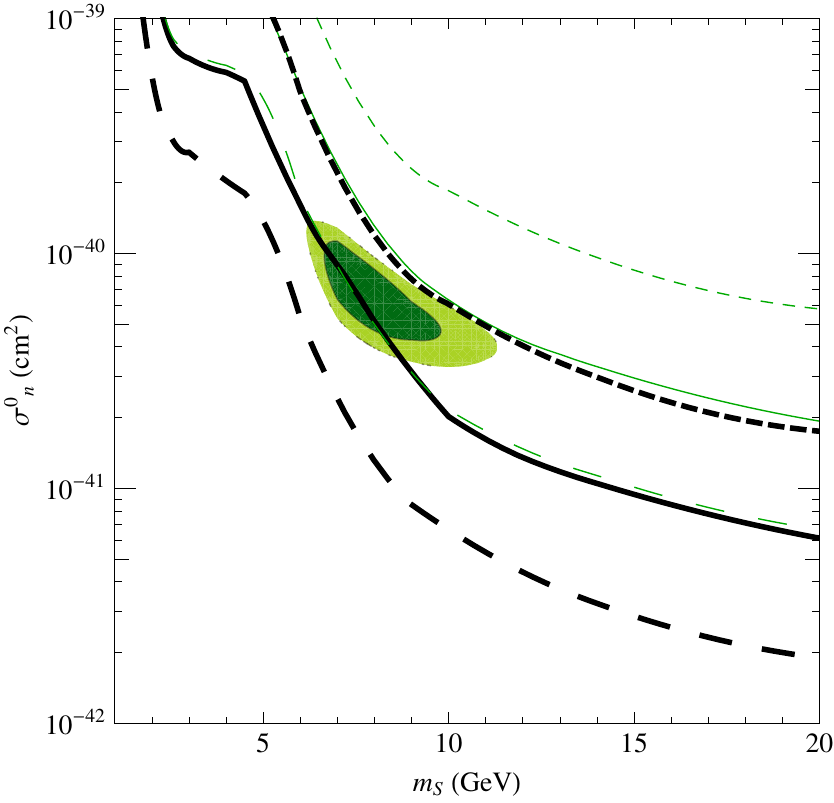}
\end{minipage}
\caption{Left panel: the SI cross-section ($\sigma^0_{n}$) {\em vs} the scalar singlet mass ($M_S$) with CoGeNT (the region in the middle, in green), and DAMA (the regions in purple, above without channelling and below with channelling). The contours are given for 90 and 99.9 \% C.L. The single continuous line (in red) is the exclusion limit from Xenon10 (95 \% C.L.). The dashed (blue) line corresponds to the CDMS-Si limit. The region between the two continuous black line is the one corresponding to $S$ candidates with WMAP relic abundance.
Right panel: only the CoGeNT region, together with 95 \% C.L. exclusion limits from isotropic diffuse gamma-ray emission observation by {\it Fermi}-LAT. The lines correspond to distinct astrophysical assumptions (see \cite{Arina:2010rb} for details). The region between the two continuous (in black and in green) lines may be considered to give  conservative (range of) exclusions limits  (see also \cite{Abazajian:2010zb}).
}
\label{fig:singletSI}
\end{figure}

\bigskip
 Here we have briefly reviewed the constraints on a light singlet scalar dark matter candidate. There is currently a fascinating interplay between direct detection experiments and indirect detection which are simultaneously becoming sensitive to WIMP candidates with a  mass in the GeV range. This is of course particularly relevant in the light of the recent hints for the existence of a light DM. Further relevant constraints may come from colliders \cite{Goodman:2010yf,Bai:2010hh}, as well as indirect searches, in particular antiprotons \cite{Bringmann:2009ca,Lavalle:2010yw,Low:2010tu} and synchroton radiation \cite{Bringmann:2009ca,Boehm:2010kg,Hooper:2010im}.  Specifically, current radio observations severely constrain the annihilation of  GeV candidates at the centre of the Galaxy \cite{Bringmann:2009ca,Boehm:2010kg}.

\section*{Acknowledgments}
This talk is a summary of  results obtained in works done in collaboration with S. Andreas, C. Arina, Th. Hambye and F.S. Ling. My work is supported by the FNRS-FRS, the IISN and the Belgian Science Policy (IAP VI-11).

\bibliography{bibliography_chiara}

%

\end{document}